\documentclass[a4paper]{aa} 
\usepackage{graphicx}
\usepackage{natbib}
\def\cuad{\hbox{$\,\,\!\!^{\rm 2}$}}
\usepackage{txfonts}
\begin{document}
   \title{Mapping the circumstellar SiO maser emission in R\,Leo}

   \author{R. Soria-Ruiz\inst{1}  \and J. Alcolea\inst{2} \and F. Colomer\inst{3} 
   \and V. Bujarrabal\inst{3} \and J.-F. Desmurs\inst{2} 
    }

   \offprints{{R. Soria-Ruiz}\\ \email{soria@jive.nl}}

   \institute{Joint Institute for VLBI in Europe, Postbus 2, 7990 AA Dwingeloo, The Netherlands \and
   Observatorio Astron\'omico Nacional, Alfonso XII 3, E-28014 Madrid, Spain \and 
   Observatorio Astron\'omico Nacional, Apartado 112, E-28803 Alcal\'a de Henares, Spain  
    }
   
   \date{Received ***, 2006; accepted ***, 2006}
  
  \abstract
{The study of the innermost circumstellar layers around AGB stars is crucial to understand 
how these envelopes are formed and evolve. The SiO maser emission occurs at a few stellar
radii from the central star, providing direct information on the stellar pulsation and on the
chemical and physical properties of these regions. Our data also shed light on several aspects
of the SiO maser pumping theory that are not well understood yet.}
{We aim to determine the relative spatial distribution of the 43\,GHz and 86\,GHz SiO
maser lines in the oxygen-rich evolved star R\,Leo. }
{We have imaged with milliarcsecond resolution, by means of Very Long Baseline
Interferometry,
the 43\,GHz ($^{28}$SiO $v$=1, 2 $J$=1--0 and $^{29}$SiO $v$=0 $J$=1--0) and 86\,GHz
($^{28}$SiO $v$=1 $J$=2--1 and $^{29}$SiO $v$=0 $J$=2--1) masing regions.}
  {We confirm previous results obtained in other oxygen-rich envelopes. In particular, when
comparing the 43\,GHz emitting regions, the $^{28}$SiO $v$=2 $J$=1--0 transition is produced in an
inner layer, slightly closer to the central star than the $v$=1 $J$=1--0. On the other hand, the
86\,GHz $^{28}$SiO $v$=1 $J$=2--1 line arises in a clearly farther shell. We have also mapped for
the first time the $^{29}$SiO $v$=0 $J$=1--0 emission in R\,Leo. The already reported discrepancy
between the observed distributions of the different maser lines and the theoretical predictions is
also found in R\,Leo.}
  {}
  \maketitle


\section{Introduction}

Along the Asymptotic Giant Branch phase, many stars exhibit maser amplification in different
molecular lines. In oxygen rich stars, [O]/[C]$>$1, O-bearing compounds are mainly formed
and maser emission is presented in SiO, H$_{2}$O and OH.  
The study of these different molecules provides information of the overall envelope, from the
inner layers dominated by the stellar pulsation (SiO masers) to the outermost regions where the
circumstellar material is expanding at constant velocity (OH masers). 

The very long baseline interferometry is a unique technique to study the compact
and very bright SiO emission, and therefore, it is particularly helpful in understanding the
different and complex processes occuring in these inner regions of the envelope. On the other hand,
the current models of pumping, either collisional or radiative, do not reproduce some
characteristics of the SiO masers that have been observed, as for example, their relative location
in the envelope.
To test these models and constrain the physical parameters of these inner shells better,
it is very useful to perform simultaneous observations of several maser transitions. For this
reason, we have carried out multi-line and multi-epoch observations in a sample of AGB stars. We
present in this paper the latest results for the Mira-type variable R\,Leo.

\section{VLBA observations}

The observations were made with the NRAO\footnote{The National Radio Astronomy
Observatory  is a facility of the National Science Foundation operated under cooperative
agreement  by Associated Universities, Inc.} Very Long Baseline Array (VLBA) on 2002
december 7.  Nearly simultaneous observations of different 43\,GHz and 86\,GHz
$^{28}$SiO and $^{29}$SiO maser transitions were performed in the variable
star R\,Leo. The 86\,GHz lines were observed in between the two 43\,GHz scans, and, therefore, we
can assume for our purpouses that the observations were simultaneous.

The data correlation was done at the VLBA correlator located in Socorro (New Mexico). 
Left and right circular polarizations (LCP \& RCP) were measured for the $^{28}$SiO $v$=1 and
$v$=2 $J$=1--0 lines, whereas only the LCP was observed in the other transitions. Since no
significant difference was found between the maps, less than 5\%, the final image is the average of
both polarizations.

\begin{table}[t]
\caption{Observed maser transitions and results of the fits.}
\begin{tabular}{c@{\,}cc@{\,}c@{\,}cc}
\cline{3-1}\cline{4-1}
\hline
\hline
&&&&\\[-9pt]
specie & transition  & restoring & $\bar{R}$ & $\triangle{R}$ & center \\
           &                  &  beam ($mas$\cuad) &($mas$)&($mas$)
&($X_{\rm c}$,$Y_{\rm c}$)\\[2pt]
\hline\hline
&&&&\\[-9pt]
 $^{28}$SiO& $v$=1 \,\, $J$=1--0& 0.78$\times$0.50 & 29.24& 6.42&
(26.9,--21.3)\\
          & $v$=1 \,\, $J$=2--1          & 0.50$\times$0.50 & 33.84&4.20
&(--35.2,--4.7) \\
	   & $v$=2 \,\, $J$=1--0          &  0.50$\times$0.50   &25.92 &6.88&
(25.5,--16.3) \\[3pt]
 $^{29}$SiO& $v$=0 \,\, $J$=1--0&  0.78$\times$0.22 & one spot &---&---\\
          & $v$=0 \,\, $J$=2--1          &  non--det. & --- &--- &---\\
\hline
\hline
\end{tabular}
\label{tab1}
\end{table}

Standard procedures for spectral line VLBI data reduction were followed in the calibration and
production of the maps. The amplitude calibration was done using the system temperatures and
antenna gain corrections for  the 86\,GHz and $^{29}$SiO data, and the template method  
for the other 43\,GHz data. The phase errors were removed in a two-step process: first,
the single-band delay corrections were derived from the continuum calibrators, OJ287 and 3C273.
Second, the fringe-rates were estimated by selecting a bright and simple-structured channel;
the corrections found were subsequently applied to the maser source. 
The maps were produced using the CLEAN deconvolution algorithm. 


 \section{Results and Data Fits}
 
\begin{figure*}[!ht]
\centering
\includegraphics[angle=-90, width=0.92\textwidth]{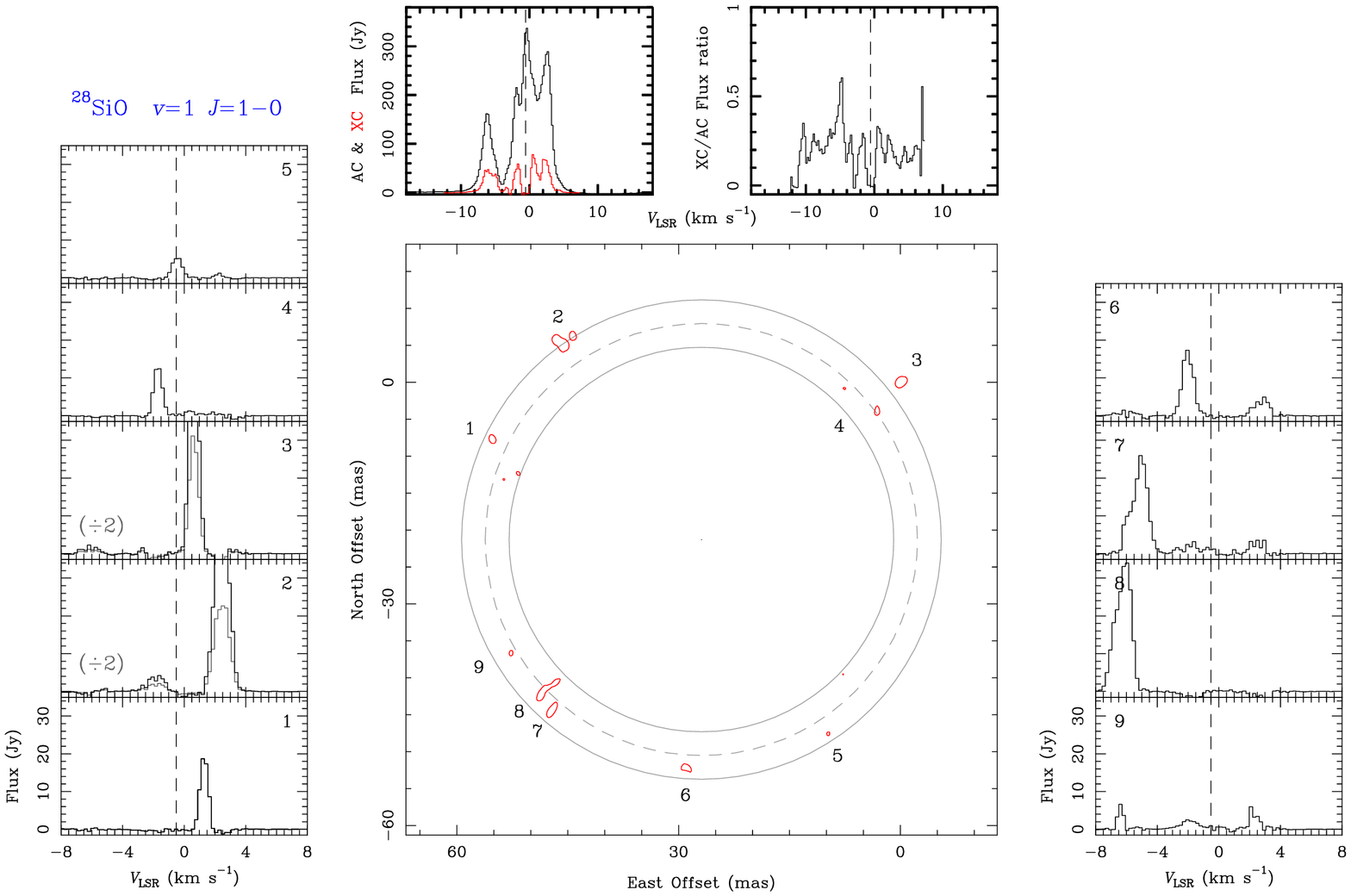}

\vspace{0.3cm}
\includegraphics[angle=-90, width=0.92\textwidth]{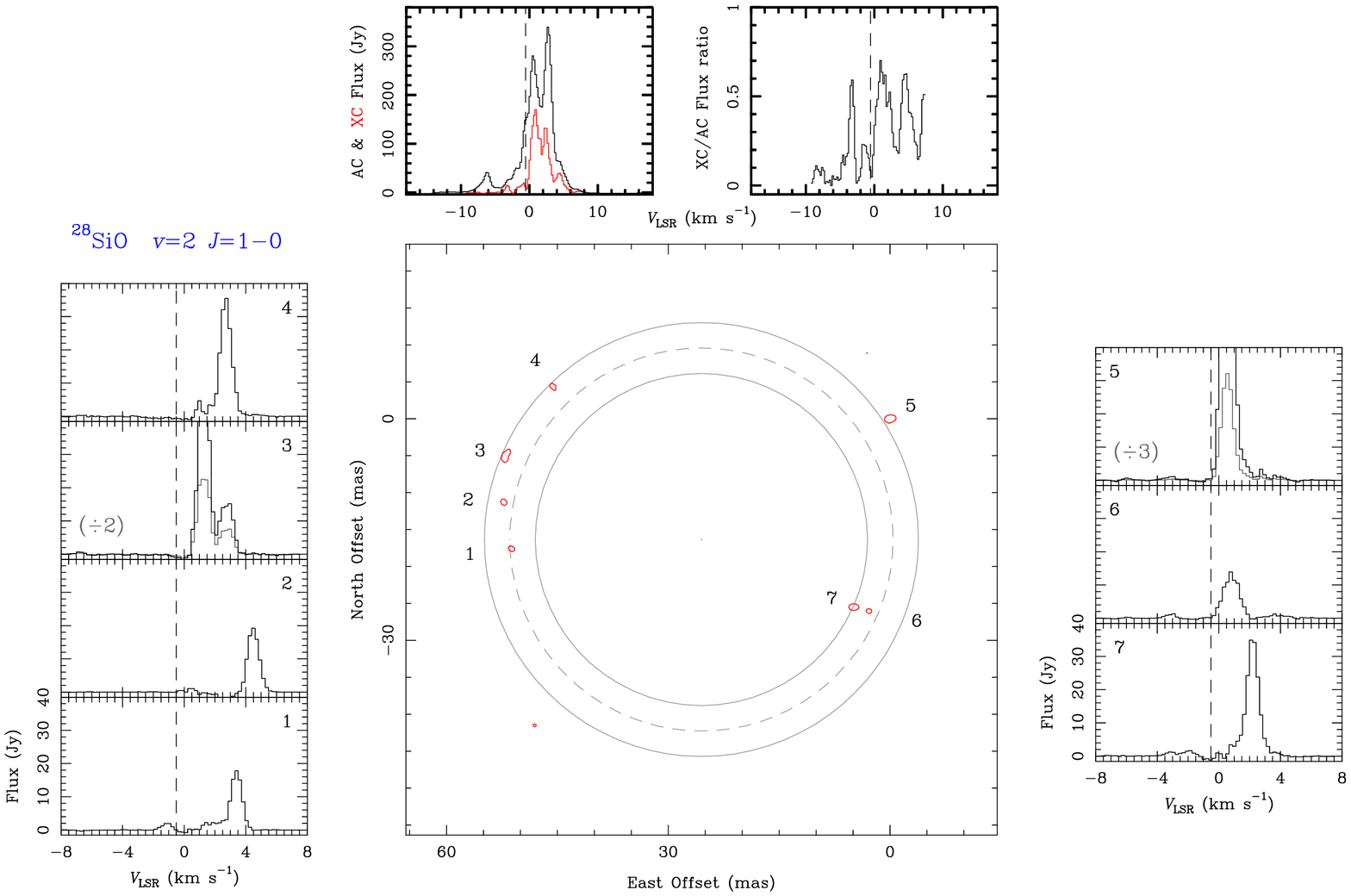}
\caption{The $^{28}$SiO $v$=1 (upper panel) and $v$=2 (lower panel) $J$=1--0 
maser emission in R\,Leo. Each figure shows the integrated intensity map in
Jy\,beam$^{-1}$\,km\,s$^{-1}$ units, the spectra of the individual maser components, the
total power spectrum and the emission in the map, and their ratio. For some maser
components, the intensity has been divided by a factor of 2 or 3 to ease the comparison with
the other spectra. The vertical dashed lines indicate the systemic velocity of the source,
$V\rm{_{LSR}}$=\,--0.5
km\,s$^{-1}$ \citep{buj89}. Circles represent the fits for the masing regions
(dashed: mean radius
$\bar{R}$, continuous: $R_{\rm{out}}$ and $R_{\rm{in}}$ defined as $\bar{R}\pm
\frac{1}{2}\triangle{R}$) (see Section 3).
The peak intensity is \mbox{28.45 Jy\,beam$^{-1}$\,km\,s$^{-1}$} ($v$=1) and 53.5
Jy\,beam$^{-1}$\,km\,s$^{-1}$ ($v$=2), and the shown contour is equivalent to the 5\,$\sigma$ level,
with $\sigma$\,=\,0.4 Jy\,beam$^{-1}$\,km\,s$^{-1}$ ($v$=1) and $\sigma$\,=\,0.7
Jy\,beam$^{-1}$\,km\,s$^{-1}$ ($v$=2).}
\label{fig1}
\end{figure*}

\begin{figure*}[!ht]
\centering

\includegraphics[angle=-90, width=0.92\textwidth]{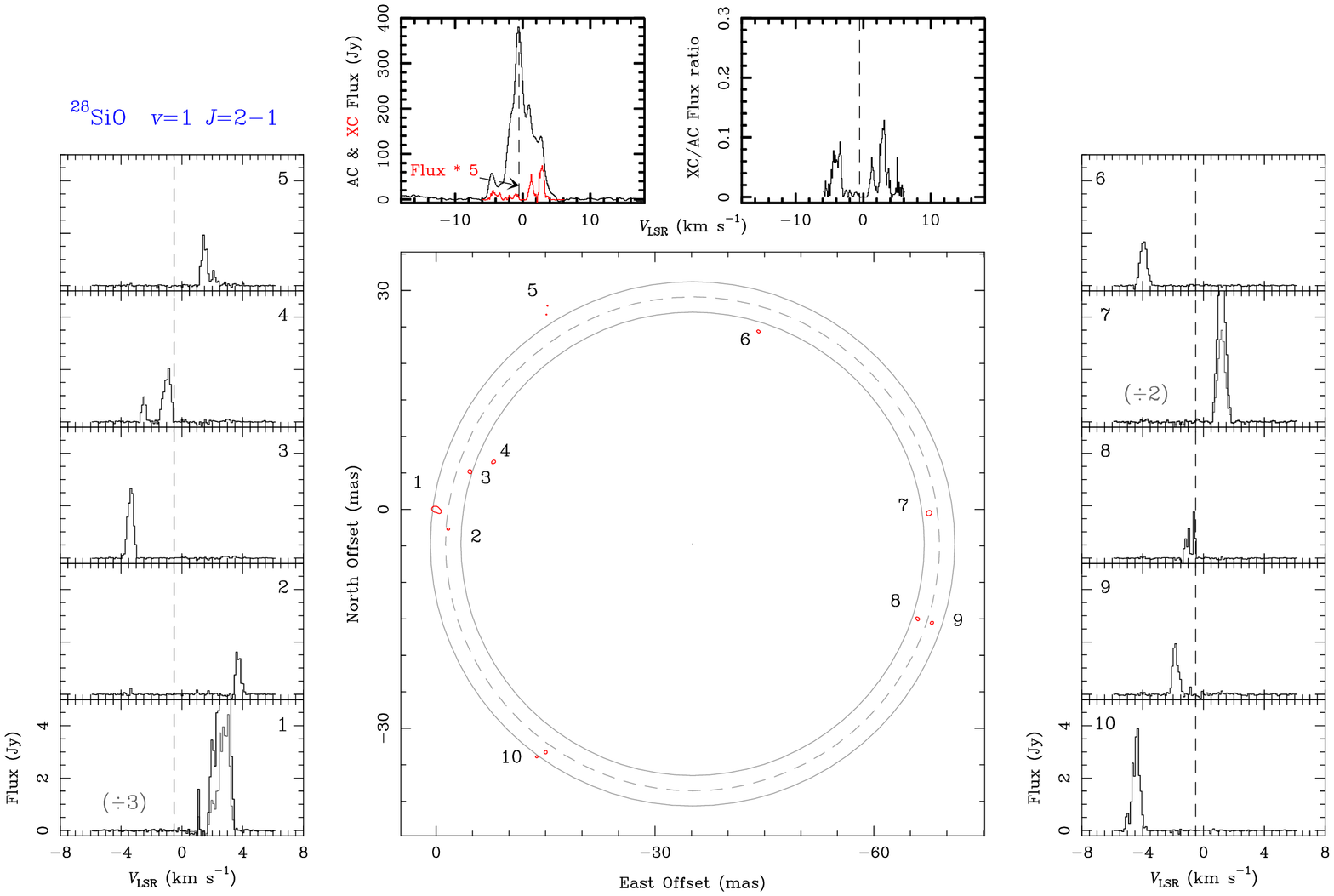}
\vspace{0.3cm}

\hspace{-4.2cm}\includegraphics[angle=-90, width=0.68\textwidth]{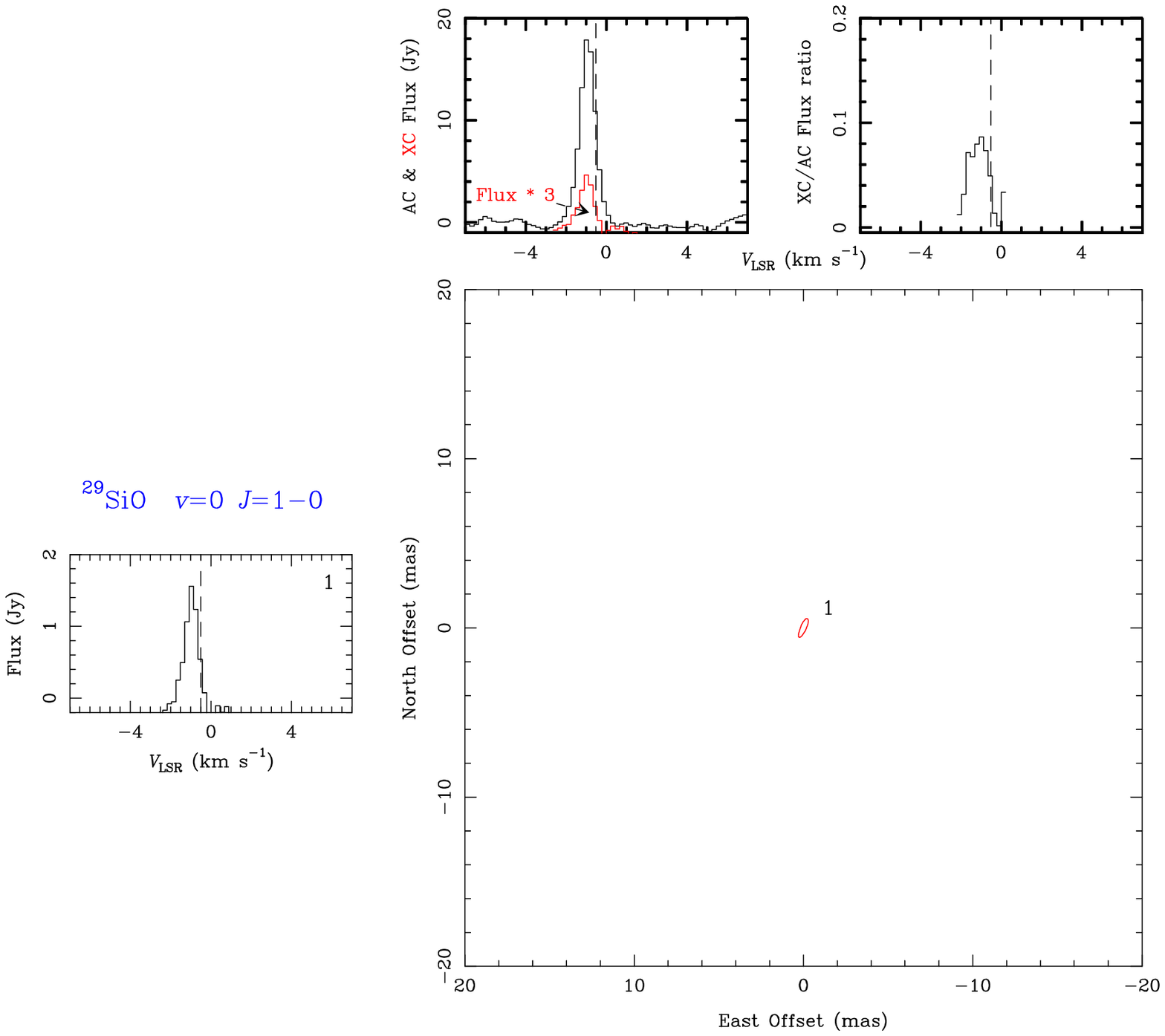}
\caption{Same as Figure \ref{fig1} for the $^{28}$SiO $v$=1 $J$=2--1 (upper panel) and $^{29}$SiO
$v$=0 $J$=1--0 (lower panel) maser emission in R\,Leo. The peak intensity is 6.03
Jy\,beam$^{-1}$\,km\,s$^{-1}$ ($^{28}$SiO) and 0.26 Jy\,beam$^{-1}$\,km\,s$^{-1}$ ($^{29}$SiO), and
the shown contour is equivalent to the 5\,$\sigma$ level, with \mbox{$\sigma$\,=\,0.08
Jy\,beam$^{-1}$\,km\,s$^{-1}$} ($^{28}$SiO) and $\sigma$\,=\,0.01 Jy\,beam$^{-1}$\,km\,s$^{-1}$
($^{29}$SiO).} 
\label{fig2}
\end{figure*}

The results are  presented as follows (Figs.\,\ref{fig1}--\ref{fig2}): for each observed
line, we show  the integrated emission map in Jy\,beam$^{-1}$\,km\,s$^{-1}$ units (center
panel),  the spectrum of the cross-correlated emission for the different maser components (numbered
panels), the total power spectrum (AC) of one of the VLBA antennas and of the emission in the map
(XC) (upper-left panel), and the ratio of these two magnitudes (upper-right panel).
We have also estimated the size of the total masing regions by fitting our data to rings. Only those
components with SNR$\geq$6 have been included in the fits. The results derived from the calculations
are summarized in Table\,\ref{tab1}: characteristic ring radius ($\bar{R}$), ring width
($\triangle{R}$) and center of the ring ($X_{\rm c}$,$Y_{\rm c}$) \citep [see
details on the fitting process in][]{soria-ruiz05}. 
In particular, the angular sizes derived for the $^{28}$SiO $v$=1 and $v$=2 $J$=1--0 
regions  (Table\,\ref{tab1}) are compatible with previous observations performed in R\,Leo by
\citet{cotton04}.
Among the six transitions observed only the $^{29}$SiO $v$=0 $J$=2--1 has not been detected.  
A more detailed description of the maps is given in the subsequent section.

\section{Relative spatial distribution and pumping mechanisms}

Our maps show that the spatial distribution of the $v$=1 and $v$=2 maser spots is similar although
not all the components appear in both transitions (Fig.\,\ref{fig1}). 
Concerning the relative location of the 43\,GHz $^{28}$SiO maser layers, the $v$=2 emission is
produced in a closer region of the envelope, assuming that the centroids of all the emissions are
coincident. This is also consistent with previously reported results in other oxygen rich envelopes
\citep[see e.g.][]{desmurs00,cotton04,soria-ruiz04,soria-ruiz05}. In contrast to the 43\,GHz
regions, this first map of the $^{28}$SiO $v$=1 $J$=2--1 emission in R\,Leo reveals that the
components of this maser line are situated in a significantly outer region of the envelope, with a
very different spot distribution. Since the $J$=2--1 emission has been imaged only in a very few
sources, this result in R\,Leo is particularly important to test the proposed SiO maser mechanisms.
Finally, the $^{29}$SiO $v$=0 \mbox{$J$=1--0} map consisted of one maser spot, thus making difficult
to derive any spatial information. The total power and recovered emission are shown in
Figure~\ref{fig2}.

Current pumping models, either radiative \citep{buj94a,buj94b} or collisional 
\citep{hum02}, predict that the different rotational maser lines within the same vibrational state
are produced under similar conditions and therefore are expected to be located in the same region of
the envelope. As previously mentioned, we find a contradiction between these theoretical
predictions and our observational results. This discrepancy has also been observed in other
oxygen-rich stars; IRC\,+10011 \citep{soria-ruiz05} and TX\,Cam \citep{SoriaEVN}. Further
calculations of the excitation of the SiO molecule in AGB stars have shown that the conditions under
which the different maser transitions occur change drastically when the line overlap between
infrared lines of H$_{2}$O and $^{28}$SiO is introduced in the pumping models
\citep{buj96,soria-ruiz04}; such a mechanism could explain the lack of coincidence between the
spots of different $J$--transitions within a vibrational state. 

Nevertheless, although these new maps support the relevance of line overlaps in the SiO maser
pumping in O--rich shells, we think that similar studies should be performed in a larger number of
evolved stars. In particular, it would be necessary to have data on all types of long-period
variable stars, namely, Mira-type, semirregular and irregular variables, as well as supergiant
stars.

\begin{acknowledgements}
This work has been financially supported by the Spanish DGI (MCYT) under projects
AYA2000-0927 and AYA2003-7584. All plots have been made using the GILDAS software package
(http://www.iram.fr/IRAMFR/GILDAS).
\end{acknowledgements}


\bibliographystyle{bibtex/aa}
\bibliography{biblio}

\begin{thebibliography}{10}
\expandafter\ifx\csname natexlab\endcsname\relax\def\natexlab#1{#1}\fi

\bibitem[{{Bujarrabal}(1994{\natexlab{a}})}]{buj94a}
{Bujarrabal}, V. 1994{\natexlab{a}}, \aap, 285, 953

\bibitem[{{Bujarrabal}(1994{\natexlab{b}})}]{buj94b}
{Bujarrabal}, V. 1994{\natexlab{b}}, \aap, 285, 971

\bibitem[{{Bujarrabal} {et~al.}(1996){Bujarrabal}, {Alcolea}, {S\'anchez
  Contreras}, \& {Colomer}}]{buj96}
{Bujarrabal}, V., {Alcolea}, J., {S\'anchez Contreras}, C., \& {Colomer}, F.
  1996, \aap, 314, 883

\bibitem[{{Bujarrabal} {et~al.}(1989){Bujarrabal}, {G\'omez-Gonz\'alez}, \&
  {Planesas}}]{buj89}
{Bujarrabal}, V., {G\'omez-Gonz\'alez}, J., \& {Planesas}, P. 1989, \aap, 219,
  256

\bibitem[{{Cotton} {et~al.}(2004){Cotton}, {Mennesson}, {Diamond}, {Perrin},
  {Coud{\' e} du Foresto}, {Chagnon}, {van Langevelde}, {Ridgway}, {Waters},
  {Vlemmings}, {Morel}, {Traub}, {Carleton}, \& {Lacasse}}]{cotton04}
{Cotton}, W.~D., {Mennesson}, B., {Diamond}, P.~J., {et~al.} 2004, \aap, 414,
  275

\bibitem[{{Desmurs} {et~al.}(2000){Desmurs}, {Bujarrabal}, {Colomer}, \&
  {Alcolea}}]{desmurs00}
{Desmurs}, J.-F., {Bujarrabal}, V., {Colomer}, F., \& {Alcolea}, J. 2000, \aap,
  360, 189

\bibitem[{{Humphreys} {et~al.}(2002){Humphreys}, {Gray}, {Yates}, {Field},
  {Bowen}, \& {Diamond}}]{hum02}
{Humphreys}, E.~M.~L., {Gray}, M.~D., {Yates}, J.~A., {et~al.} 2002, \aap, 386,
  256

\bibitem[{{Soria-Ruiz} {et~al.}(2004){Soria-Ruiz}, {Alcolea}, {Colomer},
  {Bujarrabal}, {Desmurs}, {Marvel}, \& {Diamond}}]{soria-ruiz04}
{Soria-Ruiz}, R., {Alcolea}, J., {Colomer}, F., {et~al.} 2004, \aap, 426, 131

\bibitem[{{Soria-Ruiz} {et~al.}(2005){Soria-Ruiz}, {Colomer}, {Alcolea},
  {Bujarrabal}, {Desmurs}, \& {Marvel}}]{soria-ruiz05}
{Soria-Ruiz}, R., {Colomer}, F., {Alcolea}, J., {et~al.} 2005, \aap, 432, L39

\bibitem[{{Soria-Ruiz} {et~al.}(2006){Soria-Ruiz}, {Colomer}, {Alcolea},
  {Bujarrabal}, {Desmurs}, \& {Marvel}}]{SoriaEVN}
{Soria-Ruiz}, R., {Colomer}, F., {Alcolea}, J., {et~al.} 2006, Proceedings of
  the 8$^{th}$ EVN Symposium

\end{thebibliography}
\end{document}